\documentclass[amsmath,amssymb,showpacs,twocolumn,pr]{revtex4-1}
\usepackage{graphicx,bm,color,subfigure}

\begin{document}

\title{Itinerant Ferromagnetism and $p+ip'$ Superconductivity in Doped Bilayer Silicene}

\author{Li-Da Zhang}
\affiliation {School of Physics, Beijing Institute of Technology,
Beijing 100081, China}

\author{Fan Yang}
\thanks{yangfan\_blg@bit.edu.cn}
\affiliation {School of Physics, Beijing Institute of Technology,
Beijing 100081, China}

\author{Yugui Yao}
\thanks{ygyao@bit.edu.cn}
\affiliation {School of Physics, Beijing Institute of Technology,
Beijing 100081, China}

\begin{abstract}
We study the electronic instabilities of doped bilayer silicene using the random phase approximation. In contrast to the singlet $d+id'$ superconductivity at the low doping region, we find that the system is an itinerant ferromagnet in the narrow doping regions around the Van Hove singularities, and a triplet $p+ip'$ superconductor in the vicinity of these regions. Adding a weak Kane-Mele spin-orbit coupling to the system further singles out the time-reversal invariant equal-spin helical $p+ip'$ pairing as the leading instability. The triplet pairing identified here is driven by the ferromagnetic fluctuations, which become strong and enhance the superconducting critical temperature remarkably near the phase boundaries between ferromagnetism and superconductivity.
\end{abstract}



\maketitle


\section{Introduction}

Magnetism and unconventional superconductivity (SC) as well as the intimate interplay between them have been the focuses of condensed matter physics for decades due to their rich physics and important applications. Among these subjects, the realizations of itinerant ferromagnetism (FM) and triplet SC are of particular importance in recent years. In general, the triplet SC \cite{triplet}, which is connected with topological SC \cite{tsc1,tsc2} and becomes hot topic recently, is believed to be driven by ferromagnetic spin fluctuations near the FM order. However, the realization of itinerant FM from the Stoner criterion \cite{stoner} usually requires finite and, most of the time, strong electron interactions \cite{ston1,ston2,ston3,ston4,ston5,ston6,ston7,ston8,ston9}, which is hard to deal with in the weak coupling perturbative approaches. One way to overcome this difficulty is through the introducing of divergent density of states (DOS) at the Van Hove (VH) singularities of the system, which can induce these instabilities without strong electronic interactions. It's proposed recently that, for a system with its Fermi surface (FS) doped to time-reversal (TR) variant VH saddle points, weak repulsive electron interactions can usually drive itinerant FM and triplet SC \cite{VH}.

On another front, as the Si-based counterpart of graphene, silicene has been synthesized recently \cite{syn1,syn2,syn3,syn4,syn5}, with experimental evidence showing possible SC in the doped case \cite{gap}. Furthermore, bilayer silicene (BLS) has also been available \cite{synbi}, with the energetically most favored stacking way between its two layers identified by first-principles calculations \cite{bilayer}. Based on the metallic band structure of undoped BLS, the antiferromagnetism (AFM) and the chiral $d+id'$ SC tuned by strain have been proposed \cite{bilayer}. This intriguing result motivates us to further investigate the electronic instabilities in doped BLS, specifically focusing on the VH doping levels since the divergent DOS there favors the occurrence of electronic instabilities. Paying our attention to VH doping, we notice that, in VH-doped monolayer graphene whose VH saddle points locate at TR invariant momenta, the chiral spin density wave or the chiral $d+id'$ pairing has been proposed \cite{graph1,graph2,graph3,graph4,graphsc}. Similar results have also been found in monolayer silicene \cite{silicene}. In contrast, the interesting property of the VH singularities here in BLS lies in that the VH saddle points locate at TR variant momenta. For such VH singularities, the study based on the renormalization group theory has pointed out the possibility of the formation of itinerant FM and triplet SC \cite{VH}.

In this paper, we perform the calculations based on the random phase approximation (RPA) to investigate possible electronic instabilities of doped BLS. The main results of our calculations are as follows. In addition to the $d+id'$ SC occurring at low doping levels, the itinerant FM and the triplet $p+ip'$ SC emerge as the leading instabilities of the system in the narrow doping regions around the VH singularities and the vicinity of these regions respectively. In the presence of a weak Kane-Mele spin-orbit coupling (SOC), the helical $p+ip'$ pairing becomes the leading instability of the system. The emergence of the FM and the triplet SC results from the large DOS and the strong ferromagnetic correlation near the VH singularities. Near the critical doping level separating the FM and triplet SC, the strong ferromagnetic fluctuations will greatly enhance the superconducting critical temperature, which provides possibility to realize this triplet $p+ip'$ pairing state at experimentally accessible temperatures.

The rest of this paper is organized as follows. In Sec. II, we describe the Hubbard model of BLS, as well as the RPA approach. In Sec. III, we calculate the susceptibilities of the system, and demonstrate the itinerant FM occurring around the VH singularities. In Sec. IV, we study the superconducting pairing symmetries for different doping levels, and propose that the $p+ip'$ pairing dominates over the $d+id'$ one in the vicinity of the ferromagnetic regions. Finally in Sec. V, a conclusion will be reached after discussions on the experimental detection of the novel $p+ip'$ pairing state proposed here.

\section{Model and Approach}

\begin{figure}[htbp]
\centering
\includegraphics[width=0.48\textwidth]{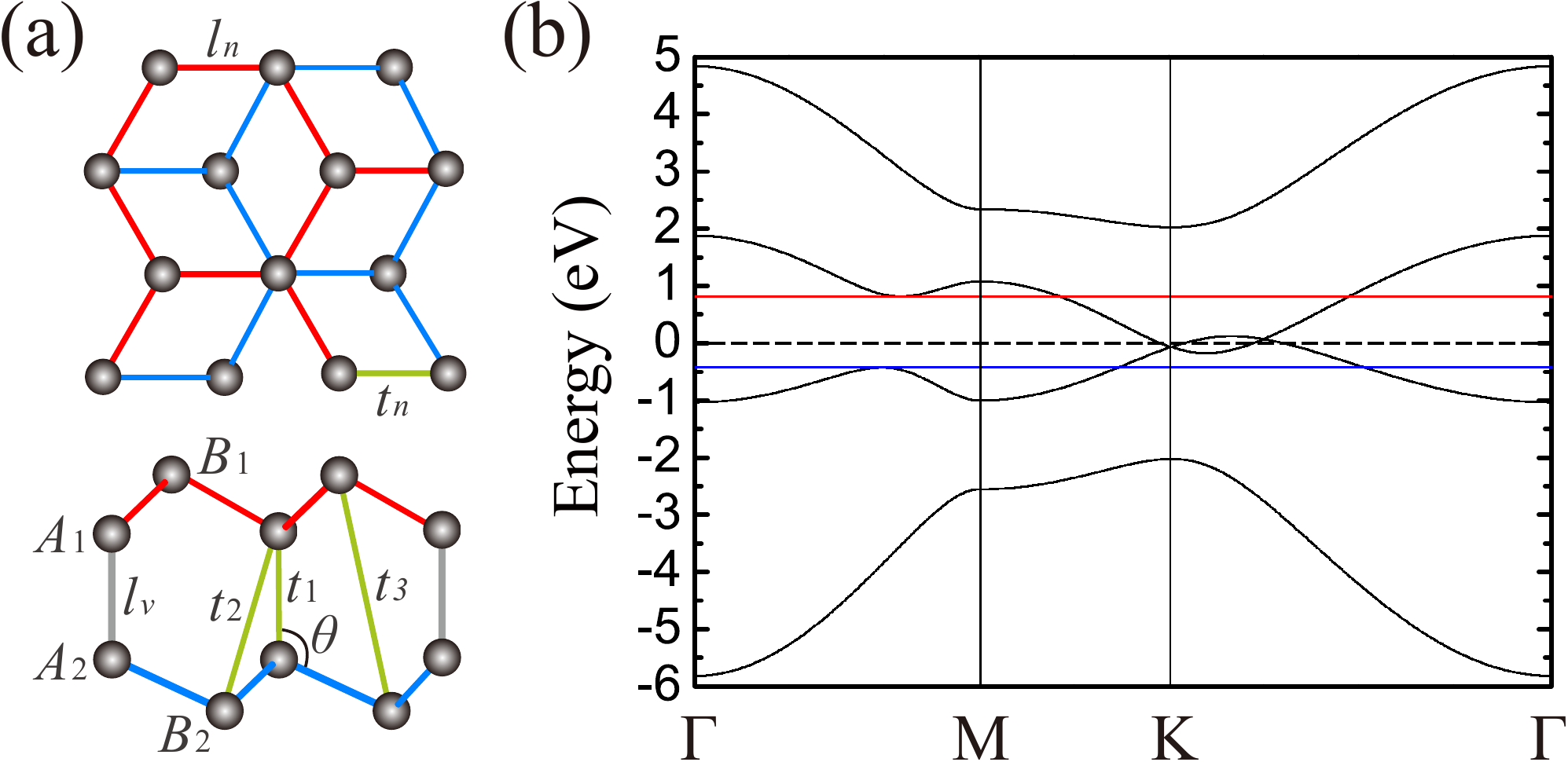}
\caption{(a) Optimized lattice structure of BLS. (b) The corresponding band structure. In (a), both the top view (upper) and side view (lower) are shown. The intralayer nearest neighbor bond length $l_n$, the vertical bond length $l_v$, and the angle $\theta$ between them are marked, together with the hopping integrals $t_n$, $t_1$, $t_2$, and $t_3$. In (b), the black dashed, red and blue solid horizontal lines denote the Fermi levels of the undoped, electron and hole VH doped systems respectively.}\label{band}
\end{figure}

The lattice structure of BLS\cite{bilayer} is shown in Fig. \ref{band}(a), which belongs to the $D_{3d}$ point group. While sublattice $A_1$ of the upper silicene layer couples vertically to sublattice $A_2$ of the lower layer with a bond-length $l_v=2.52$ {\AA}, the two sublattices $A_l$ and $B_l$ within the same layer $l$ ($=1,2$) couples to each other with a bond-length $l_n=2.32$ {\AA}. Approximately equal bond lengths, together with the bond-angle $\theta=106.65^o$ between the two bonds describes an orbital hybridization more like the $sp^3$ type than the planar $sp^2$ one. This lattice structure leads to a strong interlayer coupling, and the resulting strong bonding-antibonding splitting between orbitals $A_1$ and $A_2$ pushes them far away from the Fermi level. Thus the low energy subspace formed by orbitals $B_1$ and $B_2$ will take responsibility for the main physics of the system \cite{bilayer}. This feature of BLS is obviously different from that of bilayer graphene.

According to Ref. \cite{bilayer}, the low energy physics of BLS near the FS can be described by the following 4-band Hubbard model of the system:
\begin{align}\label{model}
H=\sum_{\bm{k}\sigma\alpha\beta}c_{\bm{k}\alpha\sigma}^
{\dagger}H_{\alpha\beta}(\bm{k})c_{\bm{k}\beta\sigma}+
U\sum_{i\alpha}n_{i\alpha\uparrow}n_{i\alpha\downarrow}.
\end{align}
Here $\sigma$, $\alpha$ ($\beta$), and $i$ denote the spin, orbital, and unit cell indices respectively, and $H(\bm{k})$ is the 4-band tight-binding (TB) Hamiltonian in the basis $\{|B_1\rangle,|B_2\rangle,|A_1\rangle,|A_2\rangle\}$. The explicit expression of the TB Hamiltonian reads \cite{bilayer}
\begin{align}\label{tb}
H(\bm{k})=\left(
\begin{array}{cccc}
\Delta  &  t_3f(\bm{k})  &  t_nf(\bm{k})^*  &  -t_2f(\bm{k})^*  \\
t_3f(\bm{k})^*  &  \Delta  &  -t_2f(\bm{k})  &  t_nf(\bm{k})  \\
t_nf(\bm{k})  &  -t_2f(\bm{k})^*  &  0  &  t_1  \\
-t_2f(\bm{k})  &  t_nf(\bm{k})^*  &  t_1  &  0  \\
\end{array}
\right).
\end{align}
Here $f(\bm{k})=\sum_{\alpha}e^{i{\bm{k}}\cdot{{\bm{R}}_{\alpha}}}$
with $\bm{R}_{\alpha}$ ($\alpha=1,2,3$) being the nearest-neighbor vector, $\Delta=-0.069$ eV is the effective on-site energy difference between atoms $A_l$ and $B_l$, the hopping integrals $t_n=1.130$ eV, $t_1=2.025$ eV, $t_2=0.152$ eV, and $t_3=0.616$ eV. Since the basis $\{|B_1\rangle,|B_2\rangle,|A_1\rangle,|A_2\rangle\}$ is mainly composed of the $3p_z$ orbital of silicon \cite{u1}, we set $U=1$ eV as a rough estimate of the Hubbard interaction.

\begin{figure}[htbp]
\centering
\includegraphics[width=0.40\textwidth]{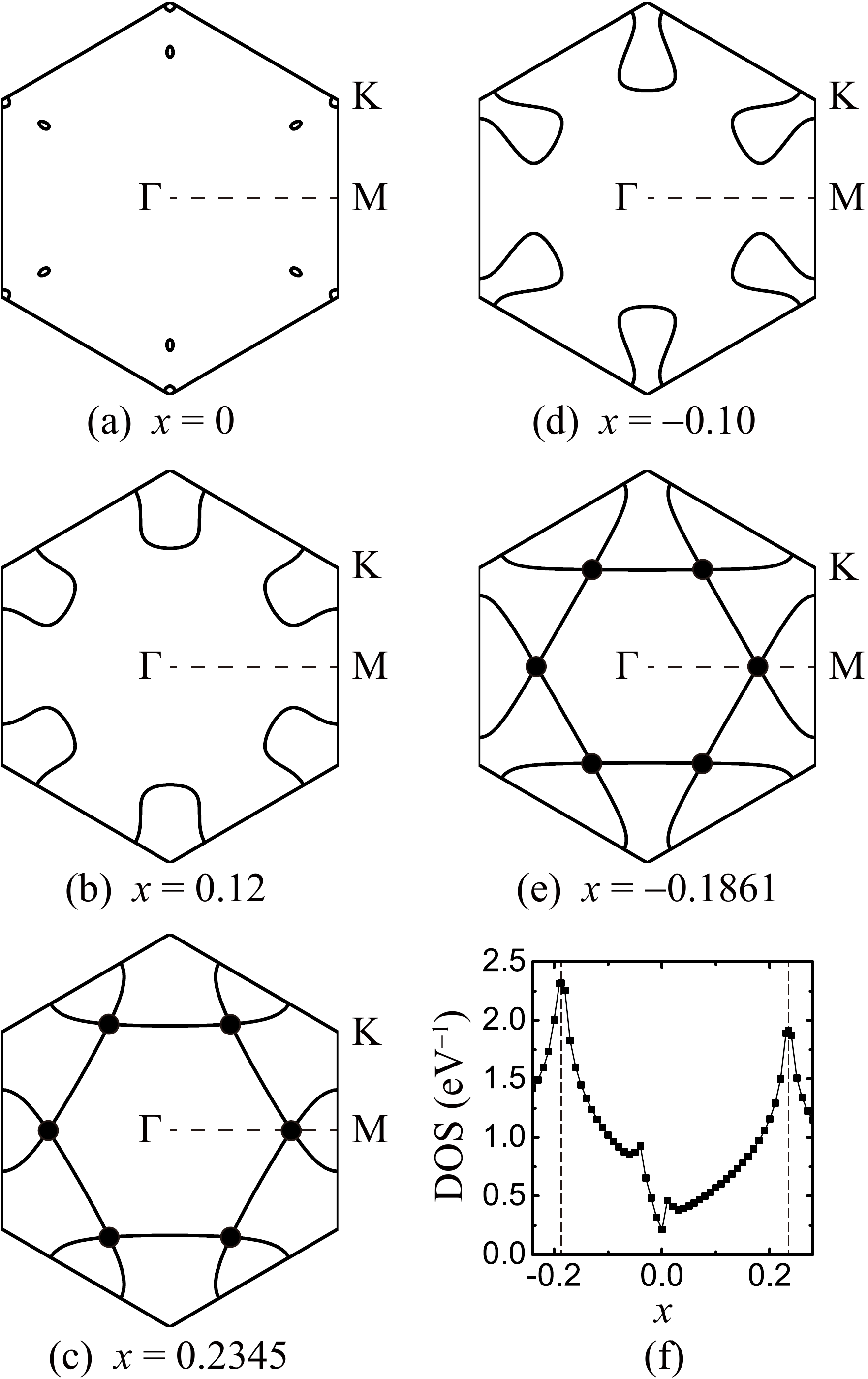}
\caption{(a)-(e) The shapes of the FS for different doping levels. The black dots in (c) and (e) indicate the VH saddle points. (f) The doping dependence of the DOS near the FS. The vertical dashed lines indicate the VH singularities.}\label{FS}
\end{figure}

The band structure for the above TB Hamiltonian (\ref{tb}) is shown in Fig. \ref{band}(b). One feature of the band structure is the $300$ meV overlap between the valence and conduction bands near the K-points. For the undoped case, this overlap causes six pairs of small electron and hole pockets around and near the K-points respectively as shown in Fig. \ref{FS}(a). The undoped system is thus intrinsically metallic and can enter a superconducting state \cite{bilayer}. When the system is doped, regardless of electron or hole doping case, the shape of the FS grows gradually from separated electron and hole pockets first to six big merged pockets around the K-points (\ref{FS}(b) and \ref{FS}(d)), which finally connect to one another at the VH saddle points, causing the Lifshits transition of the FS (\ref{FS}(c) and \ref{FS}(e)). Defining the doping level by $x=n_e-1$ where $n_e$ is the number of electrons per site, we find the doping level for the VH singularities are $x=0.2345$ for electron doping and $x=-0.1861$ for hole doping respectively. The Fermi level for these VH dopings are marked in Fig. \ref{band}(b), where the flatness of the band structure on the FS leads to the logarithmically divergent DOS near the Fermi level, as shown in Fig. \ref{FS}(f). We shall focus on these VH dopings in the following study because the divergent DOS around there urges the formation of itinerant FM and the resulting strong ferromagnetic fluctuations in the vicinity of the FM region will induce high-temperature triplet SC.

A special feature of the VH singularities of this material lies in that its VH saddle points locate on the M-$\Gamma$ axes rather than at the TR invariant M-points as in bilayer graphene as well as monolayer graphene or silicene. Such TR variant VH saddle points are named as ``type-II" VH saddle points in Ref. \cite{VH}, in contrast to the TR invariant ``type-I" VH saddle points. The ``type-II" VH singularities is special in that it allows for the formation of triplet SC. If the FS of a system contains TR invariant ``type-I" VH saddle points, the triplet pairing will not be energetically favored because its odd parity gap function will have nodes at these TR invariant VH momenta, which is no good for the energy gain. On the contrary, the TR variant ``type-II" VH saddle points of BLS locating on the M-$\Gamma$ axes provide the possibility for the system to enter the triplet pairing state.

To study the electron instabilities of this system represented by the Hubbard model (\ref{model}), we adopt the standard multi-orbital RPA approach \cite{bilayer,rpa1,rpa2,rpa3,rpa4,rpa5}. We first define and calculate the bare susceptibility tensor $\chi^{(0)l_{1},l_{2}}_{l_{3},l_{4}}\left(\bm{q},\tau\right)$. After that, the renormalized spin(s) or charge(c) susceptibilities $\chi^{(s,c)l_{1},l_{2}}_{l_{3},l_{4}}\left(\bm{q},\tau\right)$ are obtained in the RPA level. For each doping level, there will be a critical interaction strength $U_c$. For repulsive $U>U_c$, the renormalized spin susceptibility diverges, implying the formation of long-range magnetic order.  For $U<U_c$, through exchanging the spin or charge fluctuations, we obtain the effective pairing potential $V^{\alpha\beta}(\bm{k},\bm{q})$. Solving the linearized gap equation for $V^{\alpha\beta}(\bm{k},\bm{q})$ as an eigenvalue problem, we obtain the leading pairing gap function as the eigenvector corresponding to the largest eigenvalue.

\section{Itinerant Ferromagnetism}

The bare ($U=0$) susceptibility tensor of the model is defined as
\begin{align}
\chi^{(0)pq}_{st}(\bm{k},\tau)\equiv
\frac{1}{N}\sum_{\bm{k}_1\bm{k}_2}&\left\langle
T_{\tau}c_{p}^{\dagger}(\bm{k}_1,\tau)
c_{q}(\bm{k}_1+\bm{k},\tau)\right.                      \nonumber\\
&\left. c_{s}^{\dagger}(\bm{k}_2+\bm{k},0)
c_{t}(\bm{k}_2,0)\right\rangle_0,
\end{align}
Here $\langle\cdots\rangle_0$ denotes the thermal average for $U=0$, $T_{\tau}$ denotes the time-ordered product, and $p,q,s,t=1,\cdots,4$ are the sublattice indices. Fourier transformed to the imaginary frequency space, the bare susceptibility can be expressed by the following explicit formulism,
\begin{align}\label{chi0e}
\chi^{(0)pq}_{st}(\bm{k},i\omega_n)
=&\frac{1}{N}\sum_{\bm{k}'\alpha\beta}
\xi^{\alpha}_{t}(\bm{k}')
\xi^{\alpha*}_{p}(\bm{k}')
\xi^{\beta}_{q}(\bm{k}'+\bm{k})                         \nonumber\\
&\xi^{\beta*}_{s}(\bm{k}'+\bm{k})
\frac{n_F(\varepsilon^{\beta}_{\bm{k}'+\bm{k}})
-n_F(\varepsilon^{\alpha}_{\bm{k}'})}
{i\omega_n+\varepsilon^{\alpha}_{\bm{k}'}
-\varepsilon^{\beta}_{\bm{k}'+\bm{k}}}.
\end{align}
Here $i\omega_n$ is the Matsubara frequency, $\alpha,\beta=1,\cdots,4$ are the band indices, $n_F$ is the Fermi distribution function, $\varepsilon^{\alpha}_{\bm{k}}$ and $\xi^{\alpha}(\bm{k})$ are the eigenvalue and eigenvector of the TB Hamiltonian (\ref{tb}). The Hermitian static susceptibility matrix is defined as $\chi^{(0)}_{p,s}(\bm{k})\equiv\chi^{(0)pp}_{ss}(\bm{k},i\omega=0)$. For each $\bm{k}$, the largest eigenvalue $\chi^{(0)}(\bm{k})$ of this matrix represents the static susceptibility of the system in the strongest channel, and the corresponding eigenvector describes the pattern of the dominant intrinsic spin correlation in a unit cell of the system.

\begin{figure}[htbp]
\centering
\includegraphics[width=0.48\textwidth]{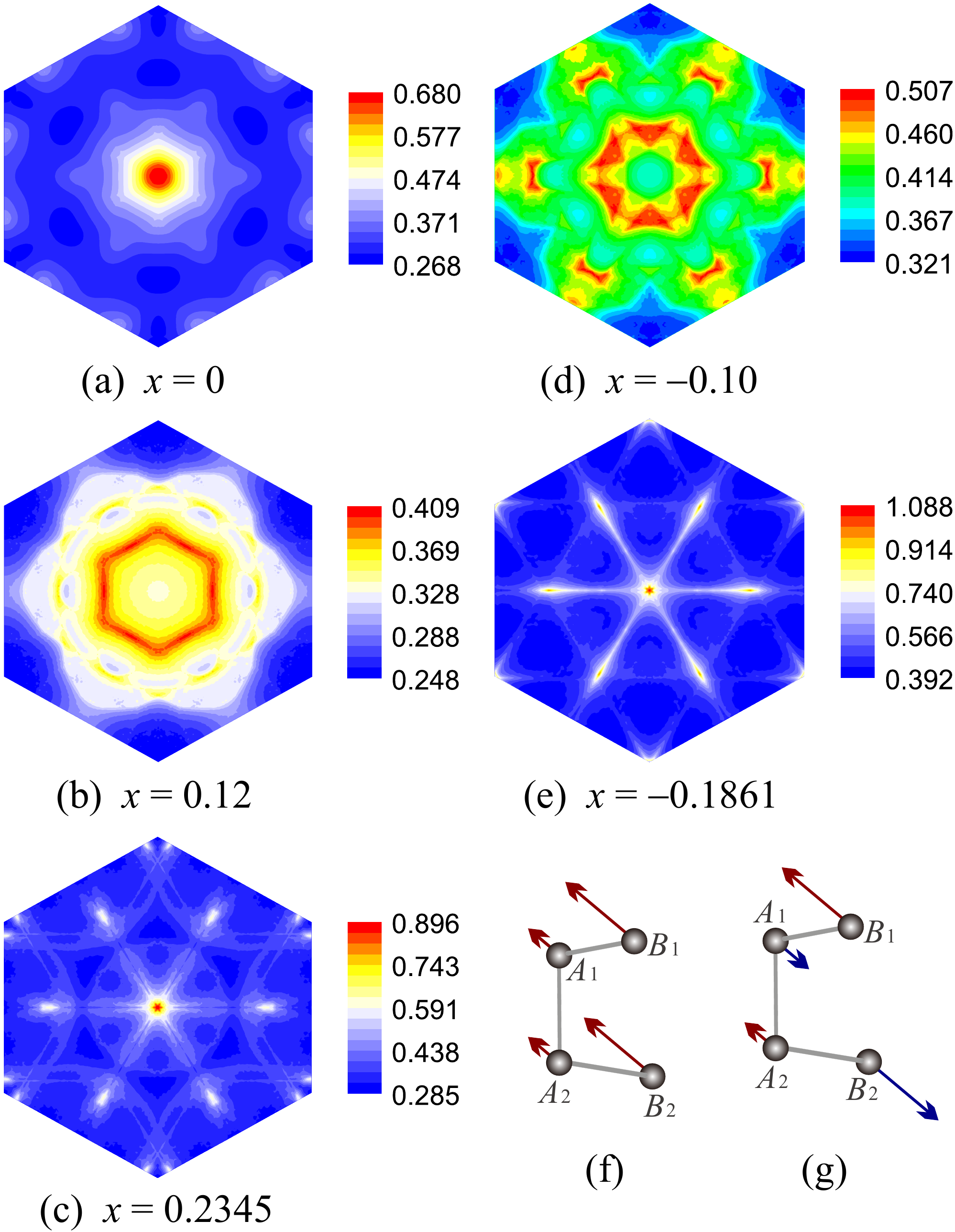}
\caption{(a)-(e) The $\bm{k}$-space distributions of the zero-temperature static susceptibility $\chi^{(0)}(\bm{k})$ for different doping levels. Typical (f) ferromagnetic pattern at the VH doping and (g) antiferromagnetic pattern at zero doping in a unit cell of the system.}\label{chi}
\end{figure}

In Figs. \ref{chi}(a)-\ref{chi}(e), we show the $\bm{k}$-space distributions of the zero-temperature static susceptibility $\chi^{(0)}(\bm{k})$ for different doping levels, which reveal the doping evolution of the static susceptibility. In particular, when the doping level changes gradually from zero to the VH doping, regardless of electron or hole doping case, the momenta of the maximum susceptibility evolve from the $\Gamma$-point (\ref{chi}(a)) first to the points around it (\ref{chi}(b) and \ref{chi}(d)), and finally back to the $\Gamma$-point again (\ref{chi}(c) and \ref{chi}(e)). Such a doping evolution of the susceptibility originates from the evolution of the FS mentioned before, and indicates that the intra-sublattice spin correlation of the system changes gradually with doping from ferromagnetic first to antiferromagnetic, and finally back to ferromagnetic again.

From the eigenvector corresponding to the largest eigenvalue of $\chi^{(0)}_{p,s}(\bm{k})$, we finds that the spin correlation within a unit cell is ferromagnetic-like (see Fig. \ref{chi}(f)) near the VH doping levels, and antiferromagnetic-like (see Fig. \ref{chi}(g)) near zero doping. Therefore, although the intra-sublattice spin correlations in both the VH doped and undoped systems are ferromagnetic, the inter-sublattice spin correlations in the former and latter cases are ferromagnetic and antiferromagnetic respectively.

When the interaction is turned on, we define the charge ($c$) and spin ($s$) susceptibilities for the model as
\begin{align}\label{chics0}
\chi^{(c)pq}_{st}&(\bm{k},\tau)\equiv\frac{1}{2N}
\sum_{\bm{k}_1\bm{k}_2\sigma_1\sigma_2}\left\langle T_{\tau}c^{\dagger}_{p\sigma_1}(\bm{k}_1,\tau)\right.               \nonumber\\
&\left. c_{q\sigma_1}(\bm{k}_1+\bm{k},\tau)
c^{\dagger}_{s\sigma_2}(\bm{k}_2+\bm{k},0)
c_{t\sigma_2}(\bm{k}_2,0)\right\rangle,                             \\
\chi^{(s^z)pq}_{st}&(\bm{k},\tau)\equiv\frac{1}{2N}
\sum_{\bm{k}_1\bm{k}_2\sigma_1\sigma_2}\sigma_1\sigma_2\left
\langle T_{\tau}c^{\dagger}_{p\sigma_1}(\bm{k}_1,\tau)\right.       \nonumber\\
&\left. c_{q\sigma_1}(\bm{k}_1+\bm{k},\tau)
c^{\dagger}_{s\sigma_2}(\bm{k}_2+\bm{k},0)
c_{t\sigma_2}(\bm{k}_2,0)\right\rangle,                             \\
\chi^{(s^{+-})pq}_{st}&(\bm{k},\tau)
\equiv\frac{1}{N}\sum_{\bm{k}_1\bm{k}_2}\left\langle T_{\tau}c^{\dagger}_{p\uparrow}(\bm{k}_1,\tau)\right.               \nonumber\\
&\left. c_{q\downarrow}(\bm{k}_1+\bm{k},\tau)
c^{\dagger}_{s\downarrow}(\bm{k}_2+\bm{k},0)
c_{t\uparrow}(\bm{k}_2,0)\right\rangle,                             \\
\chi^{(s^{-+})pq}_{st}&(\bm{k},\tau)
\equiv\frac{1}{N}\sum_{\bm{k}_1\bm{k}_2}\left\langle
T_{\tau}c^{\dagger}_{p\downarrow}(\bm{k}_1,\tau)\right.             \nonumber\\
&\left. c_{q\uparrow}(\bm{k}_1+\bm{k},\tau)
c^{\dagger}_{s\uparrow}(\bm{k}_2+\bm{k},0)
c_{t\downarrow}(\bm{k}_2,0)\right\rangle,
\end{align}
where $\sigma_1$, $\sigma_2=\uparrow$, $\downarrow$ are spin indices. For nonmagnetic states, we have $\chi^{(s^z)}=\chi^{(s^{+-})}=\chi^{(s^{-+})}\equiv\chi^{(s)}$. For $U=0$, we further have $\chi^{(c)}=\chi^{(s)}=\chi^{(0)}$.

In the standard RPA approach \cite{bilayer,rpa1,rpa2,rpa3,rpa4,rpa5}, the charge (spin) susceptibility for the model is given by
\begin{align}\label{chics}
\chi^{(c(s))}(\bm{k},i\omega_n)=
\left[I\pm\chi^{(0)}(\bm{k},i\omega_n)(U)\right]^{-1}
\chi^{(0)}(\bm{k},i\omega_n)
\end{align}
where $(U)$ is a $16\times16$ matrix, whose only four nonzero elements are $(U)^{\mu\mu}_{\mu\mu}=U$ ($\mu=1,\cdots,4$) \cite{bilayer}. Clearly, the repulsive Hubbard interaction here suppresses $\chi^{(c)}$ and enhances $\chi^{(s)}$. When the interaction parameter $U$ is weak enough, the RPA works well since all eigenvalues of the denominator matrix $\left[I\pm\chi^{(0)}(\bm{k},i\omega_n)(U)\right]$ in Eq. (\ref{chics}) are positive and hence the matrix itself has an inverse. However, if $U$ exceeds a critical value $U_c$ at which the lowest eigenvalue of $\left[I-\chi^{(0)}(\bm{k},i\omega_n)(U)\right]$ touches zero, the renormalized spin susceptibility $\chi^{(s)}$ would diverge, which implies the formation of long-range magnetic order.

\begin{figure}[htbp]
\centering
\includegraphics[width=0.48\textwidth]{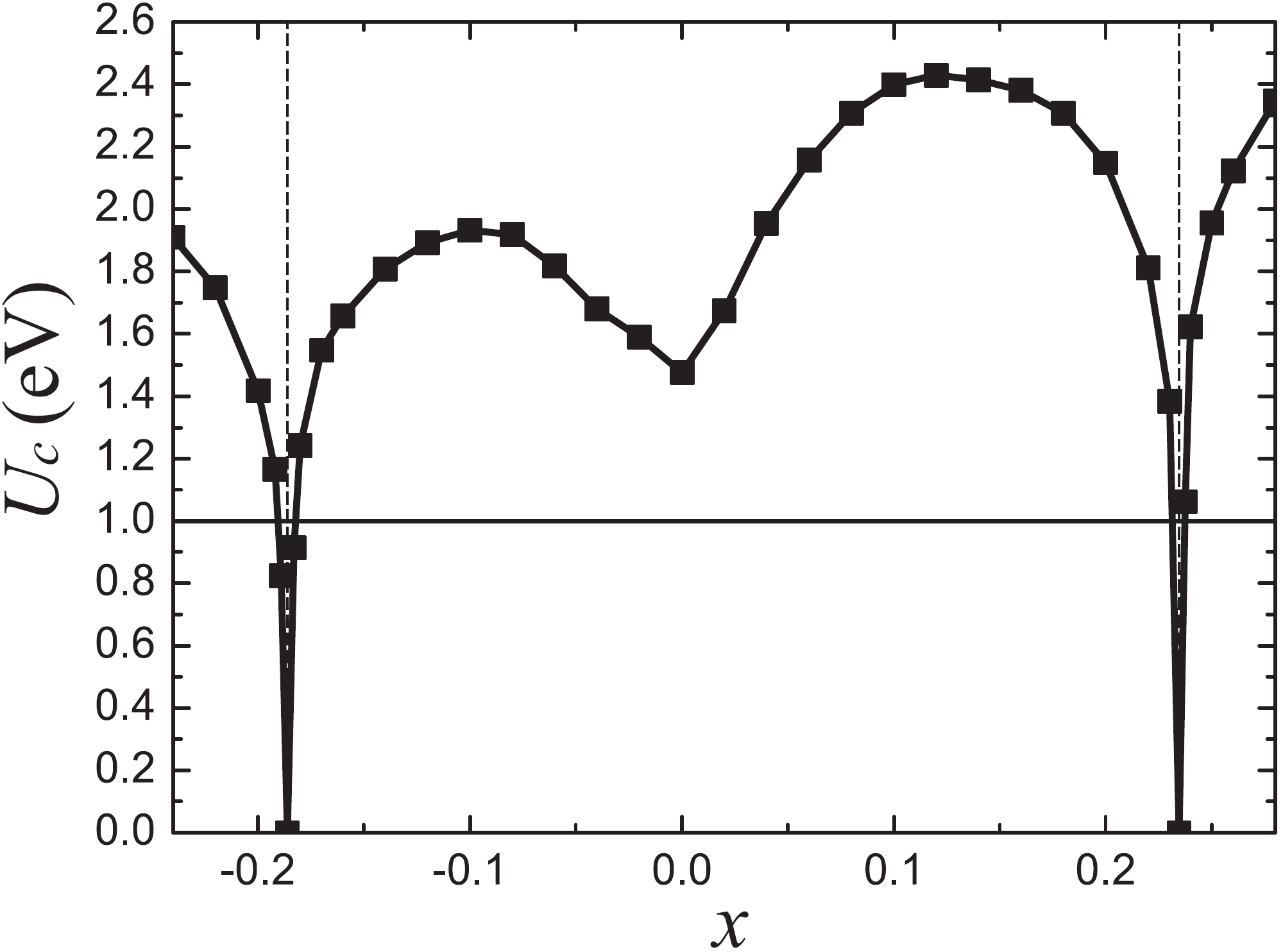}
\caption{The doping dependence of the magnetic critical interaction strength $U_c$. The horizontal solid line indicates $U=1$ eV, and the vertical dashed lines indicate the VH singularities.}\label{uc}
\end{figure}

The doping dependence of the critical interaction strength $U_c$ is shown in Fig. \ref{uc}.  The most obvious feature of Fig. \ref{uc} is that $U_c$ drops abruptly to zero near the VH singularities due to the divergent DOS there. For $U=1$ eV adopted in our calculation, the $U_c$ drops below $U$ in narrow doping regions around the VH singularities, which will lead to magnetic long-range order. From the ferromagnetic correlation near the VH singularities revealed by $\chi^{(0)}$ shown in Figs. \ref{chi}(c), \ref{chi}(e), and \ref{chi}(f), we conclude that long-range itinerant FM will emerge for $U_c<U$ in these narrow doping regions.

\section{Triplet $p+ip'$ SC}

\begin{figure}[htbp]
\centering
\includegraphics[width=0.48\textwidth]{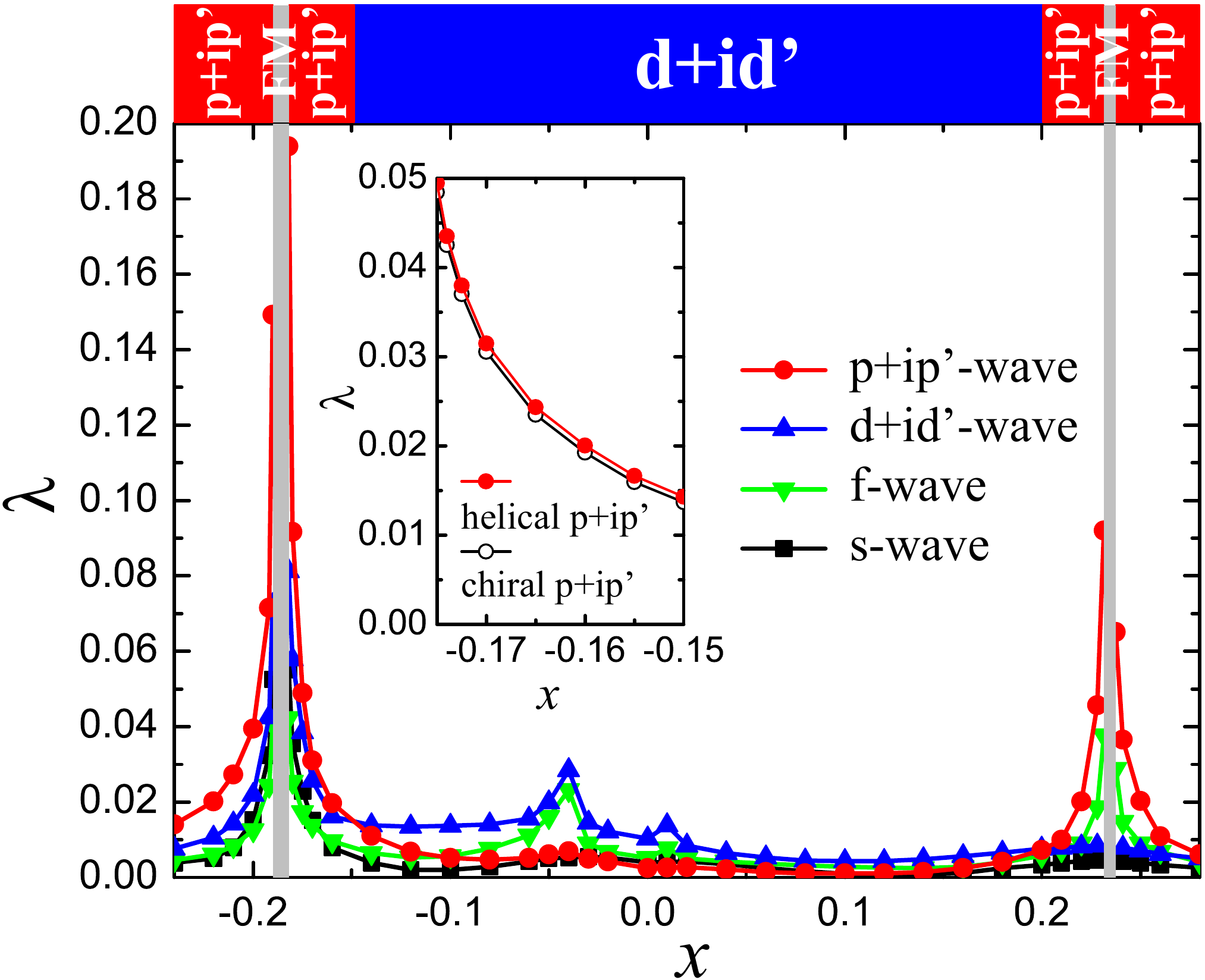}
\caption{The doping dependences of the pairing eigenvalues $\lambda$ for all possible pairing symmetries. The vertical bold grey lines indicate the doping regions where the itinerant FM occurs. Inset: the typical split between the helical $(p_x+ip_y)_{\uparrow\uparrow}, (p_x-ip_y)_{\downarrow\downarrow}$ pairing and the chiral $(p_x\pm ip_y)_{(\uparrow\downarrow+\downarrow\uparrow)}$ pairing caused by the weak SOC term with $\lambda_{SO}=10$ meV. The split in other doping regions where the $p+ip'$ SC occurs is similar to the one shown in the inset.}\label{phase}
\end{figure}

Away from the above introduced narrow doping region for itinerant FM, the interaction strength $U$ is smaller than the critical value $U_c$. Then through exchanging short-range spin or charge fluctuations between a Cooper pair, exotic superconducting states will emerge in the system as shown in Fig. \ref{phase}.

More specifically, we consider the scattering of a Cooper pair from the state $(\bm{k}',-\bm{k}')$ in the $\beta$-th ($\beta=1,\cdots,4$) band to the state $(\bm{k},-\bm{k})$ in the $\alpha$-th ($\alpha=1,\cdots,4$) band via exchanging spin or charge fluctuations. This scattering process leads to the following effective interaction vertex \cite{rpa5}:
\begin{align}\label{v}
V^{\alpha\beta}(\bm{k},\bm{k}')=
{\rm Re}\sum_{pqst}\Gamma^{pq}_{st}(\bm{k},\bm{k}')
\xi^{\alpha*}_{p}(\bm{k})                            \nonumber\\
\xi^{\alpha*}_{q}(-\bm{k})
\xi^{\beta}_{s}(-\bm{k}')
\xi^{\beta}_{t}(\bm{k}').
\end{align}
Here, for the singlet channel, we have
\begin{align}
\Gamma^{pq}_{st}(\bm{k},\bm{k}')
=(U)^{pt}_{qs}+&\frac{1}{4}\left[3(U)(\chi^{(s)}
-\chi^{(c)})(U)\right]^{pt}_{qs}(\bm{k}-\bm{k}')                 \nonumber\\
+&\frac{1}{4}\left[3(U)(\chi^{(s)}
-\chi^{(c)})(U)\right]^{ps}_{qt}(\bm{k}+\bm{k}'),
\end{align}
and for the triplet channel, we have
\begin{align}
\Gamma^{pq}_{st}(\bm{k},\bm{k}')
=-&\frac{1}{4}\left[(U)(\chi^{(s)}
+\chi^{(c)})(U)\right]^{pt}_{qs}(\bm{k}-\bm{k}')                 \nonumber\\
+&\frac{1}{4}\left[(U)(\chi^{(s)}
+\chi^{(c)})(U)\right]^{ps}_{qt}(\bm{k}+\bm{k}').
\end{align}
From the effective interaction vertex (\ref{v}), we obtained the following linearized gap equation \cite{rpa4} near the superconducting critical temperature $T_c$:
\begin{align}\label{gapeq}
-\frac{1}{(2\pi)^2}\sum_{\beta}\oint_{FS}dk'_{\parallel}
\frac{V^{\alpha\beta}(\bm{k},\bm{k}')}
{v^{\beta}_F(\bm{k}')}\Delta_{\beta}(\bm{k}')
=\lambda\Delta_\alpha(\bm{k}).
\end{align}
Here the integration is along various FS patches labelled by $\alpha$ or $\beta$, $v^{\beta}_F(\bm{k}')$ is the Fermi velocity and $k'_\parallel$ is the component of $\bm{k}'$ along the FS. Solving this gap equation as an eigenvalue problem, one obtains each pairing eigenvalue $\lambda$ and the corresponding normalized eigenvector $\Delta_\alpha(\bm{k})$ as the relative pairing gap function. The leading pairing symmetry is determined by the $\Delta_\alpha(\bm{k})$ corresponding to the largest $\lambda$. The critical temperature $T_c$ is determined by $\lambda$ through $T_c=\rm{cutoff~energy}\cdot e^{-1/\lambda}$, where the cutoff energy scales with the low energy bandwidth.

\begin{figure}[htbp]
\centering
\includegraphics[width=0.48\textwidth]{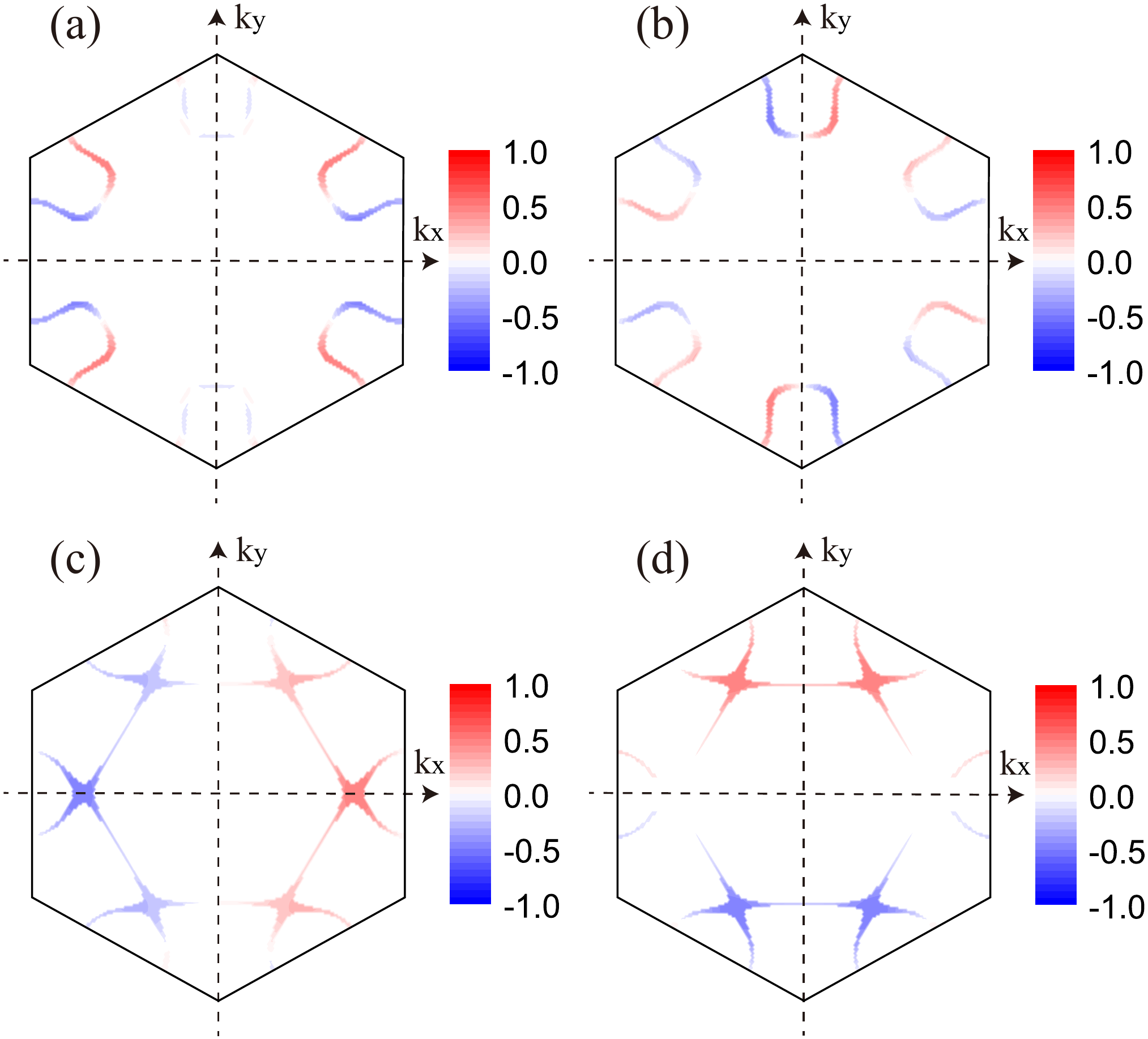}
\caption{Distributions of the gap functions on the FS: (a) $d_{x^2-y^2}$ and (b) $d_{xy}$ symmetries for doping $x=0.12$, as well as (c) $p_x$ and (d) $p_y$ symmetries for doping $x=0.24$.}\label{gap}
\end{figure}

According to its $D_{3d}$ point group, we study the possible pairing symmetries of the system including $s$, $p$, $d$, and $f$-wave ones. The doping dependences of the largest pairing eigenvalues $\lambda$ for these pairing symmetries are shown in Fig. \ref{phase}. At the low doping region, the doubly degenerate $d_{x^2-y^2}$ and $d_{xy}$ singlet pairings serve as the leading pairing symmetries, consistent with our previous results for the undoped case \cite{bilayer}. The gap function of the $d_{x^2-y^2}$ and $d_{xy}$ symmetries are symmetric and antisymmetric about the $x$-axis and $y$-axis respectively as shown in Figs. \ref{gap}(a) and \ref{gap}(b). Below $T_c$, these two degenerate pairing states will further mix to form the fully-gapped $d_{x^2-y^2}\pm id_{xy}$ (abbreviated as $d+id'$) pairings to lower the energy. Physically, the $d+id'$ singlet pairing is mediated by the antiferromagnetic spin fluctuations suggested by Figs. \ref{chi}(a), \ref{chi}(b), \ref{chi}(d), and \ref{chi}(g). More importantly, in the vicinity of the narrow doping region for itinerant FM around the VH singularities, our RPA results identify the doubly degenerate $p_x$ and $p_y$ triplet pairings as the leading pairing symmetries. The gap function of the $p_x$($p_y$) symmetry is symmetric about the $x(y)$-axis and antisymmetric about the $y(x)$-axis respectively, with gap nodes on the $y(x)$-axis, as shown in Fig. \ref{gap}(c)(\ref{gap}(d)). The emergence of the triplet $p_x$ and $p_y$ pairings near the VH singularities is the physical consequence of the strong ferromagnetic spin fluctuations there, as revealed by Figs. \ref{chi}(c), \ref{chi}(e), and \ref{chi}(f).

Since the two $p$-wave pairing symmetries are degenerate, they will probably mix to lower the energy below the critical temperature $T_c$. To determine this mixture, we set $\Delta^{\alpha}_{\bm{k}}=K_1p^{\alpha}_x(\bm{k})
+(K_2+iK_3)p^{\alpha}_y(\bm{k})$, where $p^{\alpha}_x(\bm{k})$ and $p^{\alpha}_y(\bm{k})$ denote the normalized gap functions of corresponding symmetries. Then the mixing coefficients $K_1$, $K_2$, and $K_3$ are determined by the minimization of the total mean-field energy
\begin{align}\label{ene}
E=&\sum_{\bm{k}\alpha}\varepsilon^{\alpha}_{\bm{k}}
\left[1-\frac{\varepsilon^{\alpha}_{\bm{k}}-\mu}
{\sqrt{(\varepsilon^{\alpha}_{\bm{k}}-\mu)^2
+|\Delta^{\alpha}_{\bm{k}}|^2}}\right]                               \nonumber\\
&+\frac{1}{4N}\sum_{\bm{k}\bm{k}'\alpha\beta}
V^{\alpha\beta}(\bm{k},\bm{k}')                                      \nonumber\\
&\times\frac{(\Delta^{\alpha}_{\bm{k}})^*}
{\sqrt{(\varepsilon^{\alpha}_{\bm{k}}-\mu)^2
+|\Delta^{\alpha}_{\bm{k}}|^2}}
\frac{\Delta^{\beta}_{\bm{k}'}}
{\sqrt{(\varepsilon^{\beta}_{\bm{k}'}-\mu)^2
+|\Delta^{\beta}_{\bm{k}'}|^2}}.
\end{align}
Here the chemical potential $\mu$ is determined by the constraint of the average electron number in the superconducting state. Our energy minimization gives $K_1=\pm K_3$ and $K_2=0$, which leads to the fully-gapped $p_x\pm ip_y$ (abbreviated as $p+ip'$) SC. This mixture of the two $p$-wave pairings satisfies the requirement that the gap nodes should avoid the FS to lower the energy. Note that there can be three different components of this triplet pairing with different $S_z$ quantum numbers of the Cooper pair, which are $\uparrow\uparrow,\downarrow\downarrow$, and $(\uparrow\downarrow+\downarrow\uparrow)$. In the absence of SOC, the three spin components are degenerate.

To lift up the degeneracy among the three different spin components of the triplet $p+ip'$ pairing, we add the following Kane-Mele SOC term to Hamiltonian (\ref{model}):
\begin{align}\label{H}
H_{KM}=i\lambda_{SO}\sum_{\langle\langle ij\rangle\rangle}
\nu_{ij}c_i^{\dagger}\sigma^zc_j.
\end{align}
Here $\nu_{ij}=(2/\sqrt{3})(\hat{\bm{b}}_1\times \hat{\bm{b}}_2)_z=\pm1$ with $\hat{\bm{b}}_1$ and $\hat{\bm{b}}_2$ being unit vectors along the two bonds that connect next-nearest-neighbors $i$ and $j$ on the same layer. Such a SOC term lifts up the degeneracy between the $S_z=0$ component and the $S_z=\pm 1$ components. Our RPA calculations (see the Appendix for the details) yield that the equal-spin helical $(p_x+ip_y)_{\uparrow\uparrow}$, $(p_x-ip_y)_{\downarrow\downarrow}$ pairing wins over the chiral $(p_x\pm ip_y)_{(\uparrow\downarrow+\downarrow\uparrow)}$ pairing by a small split proportional to $\lambda_{SO}$, as shown in the insets of Fig. \ref{phase} for $\lambda_{SO}=10$ meV. Such a helical triplet pairing is TR invariant weak topological SC.

It is interesting to note that the pairing eigenvalue $\lambda$ of the $p+ip'$ SC diverges in the doping region near the phase boundaries between FM and SC, due to the divergently strong ferromagnetic spin fluctuations in that region. Although the divergence of $\lambda$ is an artifact in the RPA caused by ignorance of the renormalization of the single-particle Green's function, it is possible that the strong FM fluctuations in the critical regions considerably push up the superconducting critical temperature $T_c$, which might be experimentally accessible. Taking into account that the doping level is hard to control in practice, we can instead apply tunable strain to the system to change the hopping parameters and the band structure \cite{bilayer}. As a result, the VH doping levels and the phase boundaries between FM and SC will shift so that a given doping level can access the phase boundaries to produce the high-temperature triplet $p+ip'$ SC.

\section{Discussion and Conclusion}

The unconventional triplet $p+ip'$ SC proposed here can be detected by various experiments. First of all, as an unconventional superconducting state with the phase of its pairing gap function changing on the FS, the $p+ip'$ pairing state should show no Hebel-Slichter peak upon the superconducting phase transition in the NMR relaxation rate $1/T_1T$ \cite{Hebel}. Secondly, in this triplet pairing state, the Knight shift should not obviously change below the $T_c$ \cite{Knight}. To further identify the phase structure of this pairing experimentally, we can fabricate a slice of BLS into a hexagon, and use a dc SQUID to detect the relative phase among different directions in the system \cite{squid}. In particular, determined by the $p+ip'$-wave symmetry, the phase difference between the opposite (adjacent) edges of the hexagon should be $\pi$ ($\pi/3$).

Although the $p$-wave SC is unconventional, the mixing of the $p_x$ and $p_y$ pairings into the complex $p+ip'$ one leads to a fully-gapped superconducting state, which looks similar to the conventional $s$-wave one in many aspects. For example, near zero temperature, the specific heat, the penetration depth, and the NMR relaxation rate in both fully-gapped pairing states decay exponentially with temperature. What's more, the STM spectra of both fully-gapped superconducting states should exhibit U-shaped $dI/dV-V$ curves. However, all these expected experimental results can be changed by a uniaxial strain applied on the system. More specifically, the $p+ip'$ mixing proposed here is based on the degeneracy between the $p_x$ and $p_y$ pairing states, and the degeneracy itself originates from the $D_{3d}$ point group of BLS \cite{bilayer}. Thus, by applying a uniaxial strain to break the $D_{3d}$ symmetry of the system, we can eliminate the $p+ip'$ mixing, and leave a single real $p$-wave pairing as the leading instability. Such a $p$-wave pairing can be the $p_x$ or $p_y$ one, which is determined by the axis of the applied strain. Because the resulting $p_x$ or $p_y$ pairing has gap nodes on the FS, the above-mentioned exponential temperature dependence of the experimental quantities in BLS will be replaced by power-law ones. Meanwhile, the U-shaped STM spectrum of BLS will be replaced by a V-shaped one.

In conclusion, we have systematically studied the possible electronic instabilities of doped BLS. The results of our RPA calculations predict that the system is an itinerant ferromagnet in the narrow doping regions around the VH singularities, and a triplet $p+ip'$ superconductor with a possible high $T_c$ in the vicinity of these regions. With an extra weak Kane-Mele SOC, we further single out the equal-spin helical $p+ip'$ pairing state as the leading one. This intriguing triplet superconducting state has TR invariant weak topological property, and can harbor the Majorana zero-mode at its boundary \cite{tsc2,maj1,maj2,maj3}, which is useful in the topological quantum computation.

\section*{Acknowledgements}

This work is supported by the MOST Project of China (Grants Nos. 2014CB920903, 2011CBA00100), the NSFC of China (Grant Nos. 11274041, 11174337, 11225418, 11334012), and the Specialized Research Fund for the Doctoral Program of Higher Education of China (Grants Nos. 20121101110046, 20121101120046). F.Y. is also supported by the NCET program under Grant No. NCET-12-0038.

\section{Appendix: RPA with the Kane-Mele SOC}

The Kane-Mele SOC term breaks the $SU(2)$ spin-rotation symmetry, but keeps the $U(1)$ spin-rotation symmetry around the $S_z$-axis, leaving the $S_z$-component of the total spin to be a good quantum number. Therefore, we define the following susceptibility tensors,
\begin{align}
\chi^{(1)pq}_{st}(\bm{k},\tau)\equiv
\frac{1}{N}\sum_{\bm{k}_1\bm{k}_2}&\left\langle
T_{\tau}c_{p\uparrow}^{\dagger}(\bm{k}_1,\tau)
c_{q\uparrow}(\bm{k}_1+\bm{k},\tau)\right.                        \nonumber\\
&\left. c_{s\uparrow}^{\dagger}(\bm{k}_2+\bm{k},0)
c_{t\uparrow}(\bm{k}_2,0)\right\rangle,                           \\
\chi^{(2)pq}_{st}(\bm{k},\tau)\equiv
\frac{1}{N}\sum_{\bm{k}_1\bm{k}_2}&\left\langle
T_{\tau}c_{p\uparrow}^{\dagger}(\bm{k}_1,\tau)
c_{q\uparrow}(\bm{k}_1+\bm{k},\tau)\right.                        \nonumber\\
&\left. c_{s\downarrow}^{\dagger}(\bm{k}_2+\bm{k},0)
c_{t\downarrow}(\bm{k}_2,0)\right\rangle,                         \\
\chi^{(3)pq}_{st}(\bm{k},\tau)\equiv
\frac{1}{N}\sum_{\bm{k}_1\bm{k}_2}&\left\langle
T_{\tau}c_{p\downarrow}^{\dagger}(\bm{k}_1,\tau)
c_{q\downarrow}(\bm{k}_1+\bm{k},\tau)\right.                      \nonumber\\
&\left. c_{s\uparrow}^{\dagger}(\bm{k}_2+\bm{k},0)
c_{t\uparrow}(\bm{k}_2,0)\right\rangle,                           \\
\chi^{(4)pq}_{st}(\bm{k},\tau)\equiv
\frac{1}{N}\sum_{\bm{k}_1\bm{k}_2}&\left\langle
T_{\tau}c_{p\downarrow}^{\dagger}(\bm{k}_1,\tau)
c_{q\downarrow}(\bm{k}_1+\bm{k},\tau)\right.                      \nonumber\\
&\left. c_{s\downarrow}^{\dagger}(\bm{k}_2+\bm{k},0)
c_{t\downarrow}(\bm{k}_2,0)\right\rangle,                         \\
\chi^{(5)pq}_{st}(\bm{k},\tau)\equiv
\frac{1}{N}\sum_{\bm{k}_1\bm{k}_2}&\left\langle
T_{\tau}c_{p\uparrow}^{\dagger}(\bm{k}_1,\tau)
c_{q\downarrow}(\bm{k}_1+\bm{k},\tau)\right.                      \nonumber\\
&\left. c_{s\downarrow}^{\dagger}(\bm{k}_2+\bm{k},0)
c_{t\uparrow}(\bm{k}_2,0)\right\rangle,                           \\
\chi^{(6)pq}_{st}(\bm{k},\tau)\equiv
\frac{1}{N}\sum_{\bm{k}_1\bm{k}_2}&\left\langle
T_{\tau}c_{p\downarrow}^{\dagger}(\bm{k}_1,\tau)
c_{q\uparrow}(\bm{k}_1+\bm{k},\tau)\right.                        \nonumber\\
&\left. c_{s\uparrow}^{\dagger}(\bm{k}_2+\bm{k},0)
c_{t\downarrow}(\bm{k}_2,0)\right\rangle.
\end{align}

For $U=0$, we have $\chi^{(2)(0)}=\chi^{(3)(0)}=0$ and
\begin{align}
\chi^{(1)(0)pq}_{st}(\bm{k}',i\omega_n)
&=\frac{1}{N}\sum_{\bm{k}'\alpha\beta}
\xi^{\alpha}_{t\uparrow}(\bm{k}')
\xi^{\alpha*}_{p\uparrow}(\bm{k}')
\xi^{\beta}_{q\uparrow}(\bm{k}'+\bm{k})                        \nonumber\\
&\xi^{\beta*}_{s\uparrow}(\bm{k}'+\bm{k})
\frac{n_F(\varepsilon^{\beta\uparrow}_{\bm{k}'+\bm{k}})
-n_F(\varepsilon^{\alpha\uparrow}_{\bm{k}'})}
{i\omega_n+\varepsilon^{\alpha\uparrow}_{\bm{k}'}
-\varepsilon^{\beta\uparrow}_{\bm{k}'+\bm{k}}},                \\
\chi^{(4)(0)pq}_{st}(\bm{k},i\omega_n)
&=\frac{1}{N}\sum_{\bm{k}'\alpha\beta}
\xi^{\alpha}_{t\downarrow}(\bm{k}')
\xi^{\alpha*}_{p\downarrow}(\bm{k}')
\xi^{\beta}_{q\downarrow}(\bm{k}'+\bm{k})                      \nonumber\\
&\xi^{\beta*}_{s\downarrow}(\bm{k}'+\bm{k})
\frac{n_F(\varepsilon^{\beta\downarrow}_{\bm{k}'+\bm{k}})
-n_F(\varepsilon^{\alpha\downarrow}_{\bm{k}'})}
{i\omega_n+\varepsilon^{\alpha\downarrow}_{\bm{k}'}
-\varepsilon^{\beta\downarrow}_{\bm{k}'+\bm{k}}},              \\
\chi^{(5)(0)pq}_{st}(\bm{k},i\omega_n)
&=\frac{1}{N}\sum_{\bm{k}'\alpha\beta}
\xi^{\alpha}_{t\uparrow}(\bm{k}')
\xi^{\alpha*}_{p\uparrow}(\bm{k}')
\xi^{\beta}_{q\downarrow}(\bm{k}'+\bm{k})                      \nonumber\\
&\xi^{\beta*}_{s\downarrow}(\bm{k}'+\bm{k})
\frac{n_F(\varepsilon^{\beta\downarrow}_{\bm{k}'+\bm{k}})
-n_F(\varepsilon^{\alpha\uparrow}_{\bm{k}'})}
{i\omega_n+\varepsilon^{\alpha\uparrow}_{\bm{k}'}
-\varepsilon^{\beta\downarrow}_{\bm{k}'+\bm{k}}},              \\
\chi^{(6)(0)pq}_{st}(\bm{k},i\omega_n)
&=\frac{1}{N}\sum_{\bm{k}'\alpha\beta}
\xi^{\alpha}_{t\downarrow}(\bm{k}')
\xi^{\alpha*}_{p\downarrow}(\bm{k}')
\xi^{\beta}_{q\uparrow}(\bm{k}'+\bm{k})                        \nonumber\\
&\xi^{\beta*}_{s\uparrow}(\bm{k}'+\bm{k})
\frac{n_F(\varepsilon^{\beta\uparrow}_{\bm{k}'+\bm{k}})
-n_F(\varepsilon^{\alpha\downarrow}_{\bm{k}'})}
{i\omega_n+\varepsilon^{\alpha\downarrow}_{\bm{k}'}
-\varepsilon^{\beta\uparrow}_{\bm{k}'+\bm{k}}}.
\end{align}

In the RPA, we have
\begin{align}
\begin{pmatrix}
\chi^{(1)}  \\
\chi^{(3)}  \\
\end{pmatrix}=&
\begin{pmatrix}
I  &  \chi^{(1)(0)}(U)  \\
\chi^{(4)(0)}(U)  &  I  \\
\end{pmatrix}^{-1}
\begin{pmatrix}
\chi^{(1)(0)}  \\  0  \\
\end{pmatrix},                 \label{chi13}\\
\begin{pmatrix}
\chi^{(2)}  \\
\chi^{(4)}  \\
\end{pmatrix}=&
\begin{pmatrix}
I  &  \chi^{(1)(0)}(U)  \\
\chi^{(4)(0)}(U)  &  I  \\
\end{pmatrix}^{-1}
\begin{pmatrix}
0  \\  \chi^{(4)(0)}  \\
\end{pmatrix},                 \label{chi24}\\
\chi^{(5)}=&\left[I-\chi^{(5)(0)}(U)\right]^{-1}\chi^{(5)(0)},  \label{chi5}\\
\chi^{(6)}=&\left[I-\chi^{(6)(0)}(U)\right]^{-1}\chi^{(6)(0)},  \label{chi6}
\end{align}
where $(U)$ is the same as that in Eq. (\ref{chics}).

With the above expressions of $\chi^{(1\sim6)}$, we consider the scattering of a Cooper pair from the state $(\bm{k}',-\bm{k}')$ in the $\beta$-th ($\beta=1,\cdots,4$) band to the state $(\bm{k},-\bm{k})$ in the $\alpha$-th ($\alpha=1,\cdots,4$) band. This scattering process leads to the following effective interaction vertices:
\begin{align}\label{vab}
V^{\alpha\beta}_{\uparrow\downarrow}
(\bm{k},\bm{k}')={\rm Re}\sum_{pqst}&
\Gamma^{pq\uparrow}_{st\downarrow}(\bm{k},\bm{k}')
\xi^{\alpha*}_{p\uparrow}(\bm{k})                        \nonumber\\
&\xi^{\alpha*}_{q\downarrow}(-\bm{k})
\xi^{\beta}_{s\downarrow}(-\bm{k}')
\xi^{\beta}_{t\uparrow}(\bm{k}'),                            \\
V^{\alpha\beta}_{\uparrow\uparrow}
(\bm{k},\bm{k}')={\rm Re}\sum_{pqst}&
\Gamma^{pq\uparrow}_{st\uparrow}(\bm{k},\bm{k}')
\xi^{\alpha*}_{p\uparrow}(\bm{k})                        \nonumber\\
&\xi^{\alpha*}_{q\uparrow}(-\bm{k})
\xi^{\beta}_{s\uparrow}(-\bm{k}')
\xi^{\beta}_{t\uparrow}(\bm{k}').
\end{align}
Here
\begin{align}
\Gamma^{pq\uparrow}_{st\downarrow}(\bm{k},\bm{k}')
=&(U)^{pt}_{qs}-\left[(U)\chi^{(3)}(U)\right]
^{pt}_{qs}(\bm{k}-\bm{k}')                           \nonumber\\
&+\left[(U)\chi^{(6)}(U)\right]
^{ps}_{qt}(\bm{k}+\bm{k}'),                           \\
\Gamma^{pq\uparrow}_{st\uparrow}(\bm{k},\bm{k}')
=&-\frac{1}{2}\left[(U)\chi^{(4)}(U)\right]
^{pt}_{qs}(\bm{k}-\bm{k}')                           \nonumber\\
&+\frac{1}{2}\left[(U)\chi^{(4)}(U)\right]
^{ps}_{qt}(\bm{k}+\bm{k}').
\end{align}
The inversion symmetry, together with the $U(1)$ spin-rotation symmetry of our system, enable us to symmetrize the effective interaction vertices into the following channels,
\begin{align}
V^{\alpha\beta}_{(e,0)}(\bm{k},\bm{k}')=&\frac{1}{2}\left[
V^{\alpha\beta}_{\uparrow\downarrow}(\bm{k},\bm{k}')+
V^{\alpha\beta}_{\uparrow\downarrow}(\bm{k},-\bm{k}')\right],        \label{v00}\\
V^{\alpha\beta}_{(o,0)}(\bm{k},\bm{k}')=&\frac{1}{2}\left[
V^{\alpha\beta}_{\uparrow\downarrow}(\bm{k},\bm{k}')-
V^{\alpha\beta}_{\uparrow\downarrow}(\bm{k},-\bm{k}')\right],        \label{v10}\\
V^{\alpha\beta}_{(o,\pm1)}(\bm{k},\bm{k}')=&
V^{\alpha\beta}_{\uparrow\uparrow}(\bm{k},\bm{k}'),                  \label{v11}
\end{align}
where the index $e$ is for the even parity pairing, and $o$ for the odd one. From these symmetrized effective interaction vertices, we obtained the following linearized gap equation near the superconducting critical temperature $T_c$:
\begin{align}\label{gapeqso}
-\frac{1}{(2\pi)^2}\sum_{\beta}\oint_{FS}dk'_{\parallel}
\frac{V^{\alpha\beta}_{(P,S_z)}(\bm{k},\bm{k}')}
{v^{\beta}_F(\bm{k}')}\Delta_{\beta}(\bm{k}')
=\lambda\Delta_\alpha(\bm{k}),
\end{align}
which replaces Eq. (\ref{gapeq}) to determine the $T_c$ and the leading pairing symmetry of the system in the presence of the Kane-Mele SOC.

\end{document}